# Hypergraph matching for MU-MIMO user grouping in wireless LANs

Xiaofu Ma [a,*], Qinghai Gao [b], Vuk Marojevic [a], Jeffrey H. Reed [a]

[a] *Bradley Department of Electrical and Computer Engineering, Virginia Tech, Blacksburg, VA, USA*
[b] *Qualcomm Atheros, San Jose, CA, USA*

**ABSTRACT**

This paper investigates the user grouping problem of downlink wireless local area networks (WLANs) with multi-user MIMO (MU-MIMO). Particularly, we focus on the problem of whether single user transmit beamforming (SU-TxBF) or MU-MIMO should be utilized, and how many users and which users should be in a multi-user (MU) group. We formulate the problem for maximizing the system throughput subject to the multi-user air time fairness (MU-ATF) criterion. We show that hypergraphs provide a suitable mathematical model and effective tool for finding the optimal or close to optimal solution. We show that the optimal grouping problem can be solved efficiently for the case where only SU-TxBF and 2-user MU groups are allowed in the system. For the general case, where any number of users can be assigned to groups of different sizes, we develop an efficient graph matching algorithm (GMA) based on graph theory principles. We evaluate the proposed algorithm in terms of system throughput using an 802.11ac emulator, which is created using collected channel measurements from an indoor environment and simulated channel samples for outdoor scenarios. We show that our GMA achieves at least 93% of the optimal system throughput in all considered test cases.

## 1. Introduction

It is predicted that by 2019 more than 60% of the smart phone and tablet traffic in the U.S. will be carried over wireless local area networks (WLANs) [1]. The IEEE 802.11 WLAN Task Groups are pursing of gigabit wireless communications to further increase the throughput and spectral efficiency [2]. As an enabling technology for increasing system throughput, beamforming in multiple antenna systems has been introduced in new generation WLANs, such as 802.11ac and 802.11ax [3]. In a beamforming system, transmit precoding techniques [4,5] enable a multiple antenna transmission system to steer one or multiple spatial streams to one or multiple end-users simultaneously sharing the same frequency band. In the single user transmit beamforming (SU-TxBF) mode, the transmitter focuses the energy towards one particular direction. Compared with the omnidirectional transmission, SU-TxBF results in a higher signal-to-interference-plus-noise ratio (SINR) at the target receiver, enabling higher data rate. In the multi-user MIMO (MU-MIMO) mode, the system transmits multiple spatial streams that are directed to spatially separated receivers. It has been shown theoretically that the capacity of an MU-MIMO system increases linearly with the number of transmit antennas or the number of receive antennas, whichever is lower [6,7]. The MU-MIMO system throughput improvement has also been demonstrated empirically in [8–10]. Another advantage over traditional MIMO systems is that MU-MIMO systems are beneficial even with single-antenna end user terminal, which reduces the device's size and cost.

In a typical WLAN scenario, it is the access point (AP) that directs simultaneous multi-stream transmissions to multiple stations. The challenge is that the number of mobile stations that need to be served can be much larger than the number of transmit antennas, $N_t$, at the AP. In such loaded WLAN systems, the allowed number of users per multi-user (MU) group, $N_u$, that the AP can serve simultaneously is less than the number of mobile stations in the network. Thus, selecting sets of user(s) using SU-TxBF or MU-MIMO transmission along with scheduling all these groups over successive time slots is essential for achieving high system throughput while guaranteeing user fairness.

In the past few years, the research efforts on MU-MIMO user selection for downlink transmission can be divided into user pairing when $N_u = 2$ and user grouping when $N_u \geq 3$. For user pairing, a low complexity solution is proposed in [11] for improved system throughput, but it cannot achieve maximal system throughput because it is a greedy approach. The optimal and polynomial time solution is presented in [12], where the authors analyze the user pairing for downlink MU-MIMO with zero forcing precoding and show that the problem can be solved optimally as a combinatorial optimization problem [13].

* Corresponding author.
  *E-mail address:* xfma@vt.edu (X. Ma).

For user grouping, two commonly used metrics for grouping users are estimated capacity and user orthogonality, which are both calculated based on the channel states. The orthogonality-based (or Frobenius norm-based) grouping approaches, such as [14], greedily group the most orthogonal users in one group with respect to channel-norm-related parameters and disperse highly-correlated users over different time slots. Capacity-based grouping approaches, such as [15], adopt estimated capacity as the metric to greedily group users for improving the system throughput. These user grouping strategies are designed primarily for cases where multiuser diversity offers abundant channel directions, i.e., the transmitter can find user groups with good spatial separation among users. An effective capacity-based strategy with a MAC layer design is presented in [16] that considers every MU transmission group has equal number of users. The available techniques do not provide the optimal framework to decide whether SU-TxBF or MU-MIMO should be used, and how many and which users should be assigned into an MU group. These are important concerns in real environments with unpredictable channel correlations among users. Particularly for scenarios where the channel correlation is high, flexible group sizes can be beneficial and SU-TxBF can outperform MU-MIMO. This has been demonstrated in different radio environments, especially in outdoor scenarios [17]. Since WLAN deployments in the coming years will include public and outdoor areas [18], a need arises for the WLAN APs to select beamforming modes according to the radio environment. This paper provides a framework for answering the question of how to group users for commercial WLAN scenarios. The main contributions are summarized as follows.

1. We model the user grouping problem using a hypergraph and show that the maximum hypergraph matching provides the optimal solution for choosing between SU-TxBF and MU-MIMO. We demonstrate an approach to determine how many users and which users to assign to the MU groups to maximize system throughput subject to MU-MIMO air time fairness.
2. We develop an efficient and scalable algorithm based on graph matching for solving the above problem and evaluate its performance using an 802.11ac emulator based on our collected channel measurements in an office environment and simulated outdoor conditions. The results show that the proposed algorithm achieves at least 93% of the optimal system throughput, which outperforms the state-of-art algorithms, such as Zero Forcing with Selection (ZFS) [15] and Semi-orthogonal User Selection (SUS) [14].

The rest of the paper is organized as follows: Section 2 presents the problem statement with system description and derives the complexity of the exhaustive search solution. In Section 3, we model the grouping problem using a hypergraph, and show that the maximum hypergraph matching provides the optimal solution. An efficient algorithm, which is based on graph matching, is proposed for the IEEE 802.11ac system. Section 4 presents the experimental results, and Section 5 concludes the paper.

## 2. Problem context and formulation

In this section, we describe the downlink MU-MIMO transmission context (Section 2.1), and introduce MU-MIMO air time fairness (Section 2.2). The formulation of the problem with the solution space analysis is given in Section 2.3.

### 2.1. Downlink MU-MIMO transmission

We consider a downlink MU-MIMO scenario in WLAN which consists of one AP with $N_t$ antennas and $M$ mobile stations, each equipped with 1-receive antenna. The AP serves all mobile stations. We consider two transmission modes: SU-TxBF and MU-MIMO. We refer to them as *SU* and $MU_X$, where $X$ stands for the number of users in the MU group. The AP supports both *SU* and $MU_X$ transmission modes and can flexibly switch between them and between $MU_X$ and $MU_Y$ ($X \neq Y$). We denote $N_u$ to be the maximum number of users in an MU group for the considered system. Similar to the MAC protocol used in [19–21], the AP acquires the channel state information (CSI) of all active users before performing downlink MU transmission in a transmit opportunity (TXOP) period.

### 2.2. Fairness criterion

Throughput maximization and fairness consideration are two important criteria for WLAN design and deployment. Throughput maximization aims to either maximize the individual throughput of each station or the overall throughput of the network. However, such strategies can result in unfair quality of service (QoS) delivery among the mobile stations.

Fairness in a non-MU-MIMO WLAN system, such as IEEE 802.11a/b/g, can be classified into two categories: throughput fairness and air time fairness (ATF). Systems employing throughput fairness provide equal throughput for each individual station in the network. However, as observed in [22], when one or more mobile stations use a lower bit rate than the others, the aggregate throughput of all stations is drastically reduced. This is because lower-rate stations capture the channel for more time. This penalizes higher-rate stations. The extreme case is where one station's rate is so low that it occupies most of the air time. To overcome this deficiency of throughput fairness, [23] and [24] present the advantages of ATF, where each competing node receives an equal share of the wireless channel time. They demonstrate that this notion of fairness can lead to significant improvements in aggregate throughput while still guaranteeing that no node receives worse channel access than it would if all stations were using the same transmission rate. In addition, it is shown that the long-term individual throughput of a station is proportional to that station's data rate capability. Thus, ATF helps to control transmissions in such a way that it gives WLAN users a better and more predictable wireless experience. Under the concept of ATF, user fairness is achieved when each user gets equal time to transmit in the network with only SISO transmissions.

We extend ATF to MU-MIMO ATF for the downlink MU-MIMO WLAN scenario. Consider that the AP serves the stations in turns in a Round-Robin manner [25]. If time duration $T_{SU}$ is allocated for a single-user transmission and time duration $T_{MU}$ is allocated for a multi-user transmission, then $T_{MU} = T_{SU} \cdot N_{mu}$ defines MU-MIMO ATF, where $N_{mu}$ is the number of users in the multi-user transmission group. In other words, MU-ATF entails that the transmission time allocated to a $MU_{N_{mu}}$ group is $N_{mu}$ times that of an *SU*.

There are two explanations why MU-MIMO ATF is more applicable than ATF across groups. First, encouraging higher order MU transmission increases the system throughput. This is because with a larger percentage of total transmission time allocated to MU transmission, there is higher spectral efficiency and thus higher throughput. Second, if the ATF is equal across groups, the actual data rate for each user in an MU transmission may be lower than that for single user transmission. This could discourage MU-MIMO transmission. To illustrate this, consider a six-station scenario where there is one SU transmission (Station-A), one MU group with two users (Station-B, Station-C), and one MU group with three users (Station-D, Station-E, Station-F). We denote the data rate for *SU*, $MU_2$, and $MU_3$ as $r_{su}$, $r_{mu2}$, and $r_{mu3}$, where typically $r_{mu2} \leq 2 \cdot r_{su}$ and $r_{mu3} \leq 3 \cdot r_{su}$ under the same channel conditions. If the air time fairness were equal across groups, then the throughput of devices operating in MU transmission (B, C, D, E, and F, in this case) would be potentially lower than the legacy devices or MU-disabled 802.11ac devices. This discourages MU transmission and MU devices. Therefore, we consider the following multi-user air time fairness (MU-ATF) criterion: the users in a group take turns being the primary user, and ATF is then the fairness among all of the primary users in an MU group and the SU receivers. For the previous 6-station example, this air time fairness criterion would then lead to transmission period **A**, (**BC**), (**CB**), (**DEF**), (**EFD**), and (**FDE**).

## 2.3. Problem formulation and solution space

Consider $M$ single-antenna mobile stations, Station-1 to Station-$M$, are divided into $K$ groups. We denote the user index set of the $k$th group as $G_k = \{\pi_k(1), \ldots, \pi_k(|G_k|)\}$, where $\pi_k(i)$ is the index of the $i$th station in group $k$. Notice that $G_k \subseteq \{1, \ldots, M\}$, and $|G_k| \leq N_u$ for any $k$. In other words, each group contains at most $N_u$ stations being a subset of the $M$ stations in the system, where $N_u$ is the system-specific maximum group size. The estimated capacity expression is

$$R(G_k) = B \cdot \sum_{m \in G_k} log_2 \left( 1 + \frac{P_m \cdot |\mathbf{h}_m \mathbf{w}_m|^2}{N_0 + \sum_{m \in G_k, i \neq m} P_i \cdot |\mathbf{h}_m \mathbf{w}_i|^2} \right), \quad (1)$$

where $B$ is the system bandwidth, $P_m$ is the power allocated for user $m$, $\mathbf{h}_m$ is the frequency-domain channel response vector for user $m$, and $\mathbf{w}_m$ is the steering vector for user $m$. $R(G_k)$ is the estimated capacity of SU-TxBF where $|G_k| = 1$; is the estimated capacity for the MU-MIMO transmission of group $G_k$ where $|G_k| \geq 2$. Thus, the user grouping problem for throughput maximization subject to MU-ATF can be formulated as follows:

$$\text{Maximize} \quad \sum_{k=1}^{K} |G_k| \cdot R(G_k),$$

$$\text{subject to} \quad \bigcup_{k=1}^{K} G_k = \{1, \ldots, M\},$$

$$G_p \cap G_q = \emptyset \quad (\forall p \neq q, 1 \leq p \leq K, 1 \leq q \leq K),$$

where $K$ and $G_k$ (for $k = 1 \ldots K$) are the decision variables. The MU-ATF is implicitly guaranteed in the maximization expression through the $|G_k|$ term. And the constraint ensures that each user appears either as a single user transmission or in one MU group.

A straightforward way to guarantee the maximum throughput is to conduct an exhaustive search over all possible grouping candidates. There are many ways to group the users even in small networks. For example, in a network of only six stations where up to three stations can be grouped for MU transmission, (1) SU-TxBF can be used for each station, (2) stations can be grouped using three $MU_2$ groups with $C(6, 2) \cdot C(4, 2) \cdot C(2, 2) = 90$[1] combinations, (3) stations can be grouped using two $MU_3$ groups with 20 combinations, (4) the six stations can be grouped as one $SU$, one $MU_2$ and one $MU_3$ with 60 combinations, and so forth. This totals 296 combinations.

The number of possible grouping combinations increases exponentially with the number of stations. It is of order $10^{n-5}$ for $6 < n < 20$ stations and $N_u = 3$, and larger for larger MU groups. Thus, the exhaustive search is not scalable and infeasible for practical applications.

## 3. Hypergraph modeling and algorithm design

We use graph theory tools to model and solve the user grouping problem. We show that the maximum hypergraph matching provides the optimal solution for maximizing system throughput subject to MU-MIMO air time fairness when we can choose between SU-TxBF and MU-MIMO. We develop algorithms for the special case where the system supports $MU_2$ and $SU$ only and for the general case where any number of users can be grouped in an MU transmission. We show that an optimal solution of polynomial time complexity exists for the special case and propose a scalable heuristic algorithm for the general case.

[1] $C(n, k) = \frac{n!}{k!(n-k)!}$

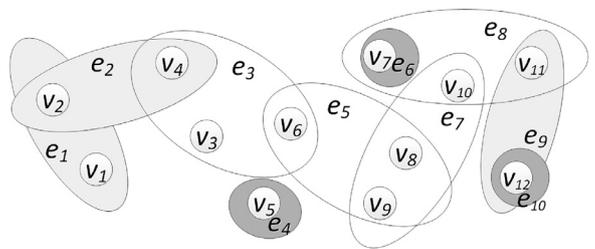

**Fig. 1.** Example of a hypergraph with twelve vertices ($v_1$ to $v_{12}$) and ten hyperedges ($e_1$ to $e_{10}$).

### 3.1. Hypergraph modeling

We use the hypergraph to model the grouping problem. A graph $G = (V, E)$ comprises a set $V$ of vertices together with a set $E$ of edges. Each edge is associated with two vertices. If the graph is weighted, each edge is given a numerical value as its weight. A hypergraph [26] is a generalization of a graph. It consists of vertices and hyperedges, where each hyperedge connects any non-zero number of vertices from $V$. In a weighted hypergraph, a single numerical value is associated with each hyperedge. An example hypergraph with twelve vertices and ten hyperedges is shown in Fig. 1. In this example, $e_1$, $e_2$ and $e_9$ are hyperedges that connect two vertices, hyperedges $e_3$, $e_5$, $e_7$ and $e_8$ connect three vertices, whereas $e_4$, $e_6$ and $e_{10}$ are each associated with a single vertex. We denote $|e_i|$ the number of vertices that $e_i$ connects.

In a hypergraph, any subset $S \subseteq E$ is called a matching if any two hyperedges in the subset do not have a common vertex, i.e., $\forall e_i, e_j \in S, e_i \cap e_j = \phi$. Subsets $\{e_1, e_3, e_4\}$, $\{e_7, e_9\}$, and $\{e_1, e_3, e_4, e_7, e_9\}$ are three example matchings for the hypergraph of Fig. 1. If every vertex in the hypergraph is incident to exactly one hyperedge of the matching, it is called a complete matching. The matching $\{e_1, e_3, e_4, e_6, e_7, e_9\}$ is a complete matching of the hypergraph example.

In our grouping problem, vertex $v_i$ in the hypergraph represents station $i$ in the network and hyperedge $e_j$ stands for an SU-TxBF or MU group. For example, $e_4$ represents a single user group with station 5, whereas $e_3$ represents an $MU_3$ group with stations 3, 4 and 6. Depending on the number of vertices that $e_j$ connects, its edge weight, $w_j$, corresponds to the estimated capacity $R(G_k)$ according to (1) for either SU-TxBF or MU-MIMO transmission of the given group (1 user, 2 users or 3 … etc.). Therefore, the problem of maximizing the throughput under the MU-ATF criterion can be reformulated as finding a complete matching $S \subseteq E$ of hyperedges such that

(1) for any two distinct hyperedges $e_i, e_j \in S, e_i \cap e_j = \phi$, and
(2) $\sum_{e_i \in S}(w_i \cdot |e_i|)$ is maximized.

The first statement ensures that any station appears in only one single group, whereas the second maximizes the sum throughput based on the MU-ATF criterion. Our system design problem can, thus, be equivalently transformed to the maximum matching problem [27] in a weighted hypergraph known from graph theory. The objective is to find a matching that maximizes the sum of the weights in the matching, which is called maximum matching in graph theory. Finding the maximum matching is a fundamental problem in combinatorial optimization and has various applications.

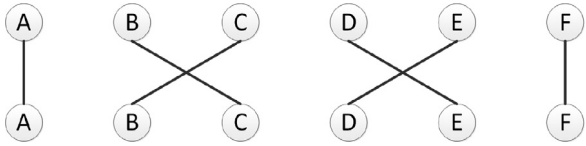

**Fig. 2.** Symmetric matching in a bipartite graph.

In a hypergraph where each hyperedge can connect any non-zero number of vertices, the maximum matching problem becomes very complex and was proven to be NP-hard [27]. All the well-known approximation algorithms for the hypergraph matching problem are based on local search methods [28]. A promising algorithm comes from Berman [29] who developed a ($\frac{k+1+\epsilon}{2}$)-approximation algorithm. There are two issues with adopting these algorithms for our application. Since all user grouping combinations are allowed, the calculation of all the corresponding hypergraph weights is proportional to $|V|^{N_u}$, which requires significant computation for reasonable problem sizes. In addition, all the local-search-based algorithms are heuristic in nature and thus do not guarantee finding the optimal solution.

Here we propose a new design approach for optimizing the sum-throughput under MU-ATF criterion while reducing the computational complexity. We start with the case where $N_u = 2$, i.e., only $MU_2$ and $SU$ are supported by the system. We then extend this to the general case where $MU_X$ (X being any positive number) and $SU$ are supported.

### 3.2. Algorithm design

We first consider the $N_u = 2$ case and show that the optimal grouping problem can be efficiently solved for systems that allow only SU-TxBF and 2-user MU groups (Section 3.2.1). We then develop an efficient graph matching algorithm based on graph theory principles for the general case where any number of users can be assigned to groups of different sizes (Section 3.2.2).

#### 3.2.1. $MU_2 + SU$

The $MU_2 + SU$ grouping problem is inherently more difficult to solve than $MU_2$ grouping. $MU_2$ user pairing [12] can be optimally solved for maximum throughput by directly adopting the Hungarian method [30]. However, this method cannot be used for solving our $MU_2 + SU$ case subject to the MU-ATF criterion. Here, we show that the grouping problem in this case can be modeled as a symmetric maximum matching problem in a weighted bipartite graph. This problem can be optimally and efficiently solved by the Blossom algorithm [31], which is a polynomial time solution.

The weighted bipartite graph is a graph whose vertices are divided into two disjoint sets $V_1$ and $V_2$. The matching then happens between vertices in $V_1$ and vertices in $V_2$. Sets $V_1$ and $V_2$ in our problem each contain all vertices in the system, where a vertex stands for a mobile station in the network. The weight of the edge connecting a station in $V_1$ with a station in $V_2$ represents the data rate of the $MU_2$ group of stations if the two connected stations are not the same; otherwise, the weight indicates for the SU-TxBF data rate for that station. Thus, the candidate matchings of the bipartite graph provide the candidate user grouping solutions, and the matching with the maximum sum weight is the optimal grouping for $MU_2 + SU$ case.

For example, in the six-station scenario, all six stations (A - F) are assigned to both sets, $V_1$ in the first row and $V_2$ in the second of Fig. 2, which illustrates the matching (**A**), (**B**C), (**C**B), (**D**E), (**E**D),

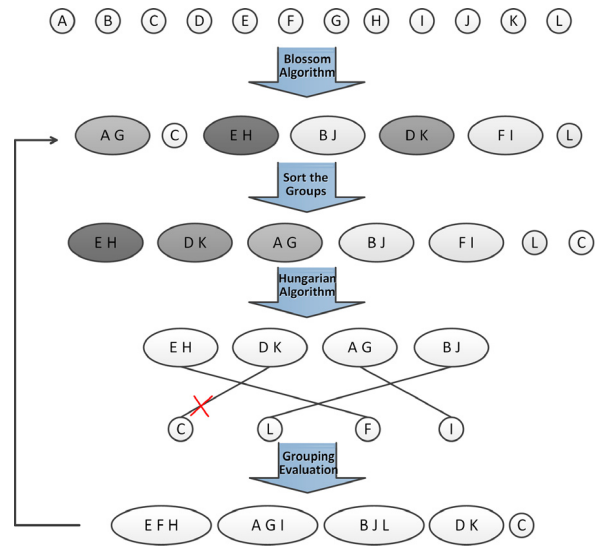

**Fig. 3.** Flowchart of the proposed graph matching algorithm (GMA).

(**F**). The weight of the edges between a station in $V_1$ and a station in $V_2$ corresponds to the data rate of the $MU_2$ group if the connected stations are different. The edge connecting the same station corresponds to SU-TxBF. For the matching of Fig. 2, A and F are SU-TxBF, and BC and DE form two MU2 groups.

Notice that the matching in Fig. 2 is a symmetric matching: if B in $V_1$ connects to C in $V_2$, then C in $V_1$ connects to B in $V_2$. This is a characteristic of our $MU_2 + SU$ problem. The symmetric matching of the weighted bipartite graph can be optimally solved using the Blossom algorithm, which has polynomial time complexity [31].

If there were no requirement on matching symmetry, B in $V_2$ could, for instance, connect to D in $V_1$ even if B in $V_1$ does not connect to D in $V_2$. In such a case where the matching is not required to be symmetric, a polynomial time solution, Hungarian algorithm [30], can be used. The Hungarian algorithm is not suitable for our $MU_2 + SU$ case, but we will use it as part of the solution for the $MU_X + SU$ case.

#### 3.2.2. $MU_x + SU$

We propose an efficient graph matching algorithm (GMA) for solving the $MU_x + SU$ problem, which directly reduces to the optimal solution for $MU_2 + SU$ case when $X = 2$. The basic design idea is to repeatedly apply weighted bipartite graph matching to form higher order MU groups and split the low data rate groups. The following steps describe the operations for obtaining the final grouping results, illustrated in Fig. 3 for a twelve-station scenario. In Step 1, the Blossom algorithm partitions the users into $MU_2$ or $SU$. These groups are then sorted in descending order of data rates in Step 2. The low throughput groups are decomposed into single user groups for the asymmetric weighted bipartite graph matching, which is solved by the Hungarian algorithm in Step 3. Notice that the purpose of the decomposition is to split the low throughput user groups as necessary for the input for Hungarian algorithm; therefore, the decomposition is not a function of any system parameters, such as network scale or user orthogonality. In Step 4, for each new group, if the group provides a higher total data rate, the grouping result is kept; otherwise, the previous group assignment remains. This procedure is executed repeatedly for $N_u > 3$. More precisely, a single run is needed for $MU_3$, two runs are needed for $MU_4$, and so forth.

This procedure is illustrated in Fig. 3 for twelve stations. The Blossom algorithm is applied to the weighted bipartite graph, grouping the station into $MU_2$ or $SU$ groups. The $MU_2$ groups here are (AG), (EH), (BJ), (DK), and (FI), whereas C and L remain as *SU* groups. The resulting groups are sorted according to their throughput and are divided into two sets, one containing the higher throughput groups and the other the lower throughput SU-TxBF users. Asymmetric weighted bipartite matching is then applied to the two disjoint sets using the Hungarian algorithm and leads to (EHF), (DKC), (AGI) and (BJL). If a newly formed group, such as (EFH), leads to a higher throughput than before, the grouping result is accepted; otherwise, the previous group assignment—(EH) and (F)—is kept. Here, the throughput of (EFH) is higher than the sum throughput of (EH) and (F). On the other hand, (DK) and (C) is a better option than (CDK). If the maximum number of users in a group is $N_u = 3$, the final grouping result is obtained as (EFH), (AGI), (BJL), (DK) and (C). If more users per MU group were permitted, the procedure would resume at Step 2. Compared with local search methods, the weight computation cost is reduced from $O(|V|^{N_u})$ to $O(N_u|V|^2)$.

**Table 1**
Pseudo-algorithm of the proposed graph matching algorithm (GMA) for $N_u$ user grouping.

| *Heuristic Algorithm for $N_u$-User Grouping* | |
|---|---|
| 1 | Calculate the achievable throughput for each single user as GMA's input; |
| 2 | Calculate the achievable $MU_2$ throughput for every two users as GMA's input; |
| 3 | Calculate the best $MU_2$ group $G$ using Blossom algorithm; |
| 4 | **for** $i = |G|:-1:1$ |
| 5 |   **if** SU transmission of the two users in $G\{i\}$ is better than $MU_2$ |
| 6 |     Add two SU groups; |
| 7 |     Remove $G\{i\}$; |
| 8 |   **end if** |
| 9 | **end for** |
| 10 | $k = 2$; |
| 11 | **while** $k < N_u$ |
| 12 |   $G_{result} = \emptyset$; $k = k + 1$; |
| 13 |   Put the groups into two sets $S_1$ and $S_2$; |
| 14 |   $S_1 = G$; $S_2 = \emptyset$; |
| 15 |   Sort the groups in $|S_1|$ according to their achievable throughput; |
| 16 |   **while** $|S_1| > |S_2|$ |
| 17 |     Put the user(s) of the lowest group of $|S_1|$ into $|S_2|$ as SU group; |
| 18 |     Remove this lowest group from $|S_1|$; |
| 19 |   **end while** |
| 20 |   **while** $|S_1| \neq |S_2|$ |
| 21 |     Put the last user of $|S_2|$ into $|S_1|$; |
| 22 |     Remove this user from $|S_2|$; |
| 23 |     **if** $|S_1| > |S_2|$ |
| 24 |       Put this SU group into $G_{result}$; |
| 25 |       Remove this user from $|S_1|$; |
| 26 |     **end if** |
| 27 |   **end while** |
| 28 |   Calculate the groups $G_{MU}$ based on Hungarian algorithm; |
| 29 |   **for** $i = |G_{MU}| : -1 : 1$ |
| 30 |     **if** the new pair achieves better throughput |
| 31 |       Add the new group into $G_{result}$; |
| 32 |     **else** |
| 33 |       Add the previous groups into $G_{result}$; |
| 34 |     **end if** |
| 35 |   **end for** |
| 36 | **end while** |

Table 1 provides the algorithm description in the form of a pseudo code. Lines 1-9 represent the use of the Blossom algorithm to group users into $MU_2 + SU$ groups. Lines 12-27 describe the sorting and splitting of the grouping results. Lines 28-35 correspond to the use of the Hungarian algorithm for forming higher order MU groups. The process then loops back to line 11 for $MU_4$ or higher-order groups. It is easy to observe that this approach provides a solution in polynomial time because the main processes are the (1) Blossom algorithm, which has the complexity of $O(|V|^3)$, (2) the sorting operation, which has a complexity of $O(|V|log|V|)$, and (3) the Hungarian algorithm, which is of complexity $O(|V|^3)$. Hence, the overall complexity order is $O(N_u|V|^3)$.

**Table 2**
System settings.

| Parameters | Values |
|---|---|
| Maximum AMPDU duration | 2 ms |
| MPDU length | 1556 bytes |
| MSDU length | 1508 bytes |
| SIFS duration | 0.016 ms |
| Bandwidth | 40 MHz |
| Number of OFDM subcarriers for data | 108 |
| Guard interval for OFDM symbol | 400 ns |

## 4. Evaluation

### 4.1. Experimental system

We obtained channel information measurements for our validation using four MIMO test nodes, each equipped with four-antennas and Qualcomm Chipset QCA9980 which is a four-spatial-stream IEEE 802.11ac transceiver chipset. We conducted the over-the-air transmission over the 40 MHz channel in an office environment. One test node operated as the transmitter AP, and the other three nodes were used to emulate twelve single-antenna receiver stations. We positioned the twelve receive antennas away from one another, at random locations in the office. The twelve sets of captured channel samples were then used to mimic the channel from one four-antenna AP to twelve single-antenna stations. We conducted the channel measurements while there was significant human movement in the measurement environment.

We implemented the measurement-driven MU-MIMO-OFDM emulator rigorously according to the IEEE 802.11ac specifications [2]. This emulator was seeded with the over-the-air channel information to test the performance of the media access control (MAC) and physical (PHY) layer algorithms. The system settings are summarized in Table 2.

To evaluate the performance of our graph matching algorithms (GMAs) proposed in Section 3, we compare it with three other state-of-the-art user grouping approaches: (a) Zero Forcing with Selection (ZFS) [15], (b) Semi-orthogonal User Selection (SUS) [14], and (c) Random Selection. ZFS selects the user with the highest channel capacity to be the first member in a group and then finds the next user based on the potential sum rate in each selection iteration. SUS picks an ungrouped user to be added to a group based on the qualified and highest effective channel norm value to the existing group. We select the best-performing SUS variant for our evaluation. Random selection randomly selects a user until the cardinality of the group reaches $N_u$. We also implement the full search (exhaustive search), which provides the optimal system throughput subject to MU-ATF. The algorithms are compared based on measured and simulated channels.

### 4.2. Evaluation based on the measured channels

Fig. 4 shows the system throughput as a function of the total number of users in the network for the system in which the largest MU transmission group size is two. We observe that the Blossom algorithm achieves the same system throughput as the full search no matter how many users exist in the network. This is because when only $MU_2$ and $SU$ are allowed in the system, the Blossom algorithm provides the optimal solution as discussed in Section 3.1. The performance of random selection is the lowest among all approaches, which is as expected because it does not take any channel information into consideration when forming transmission groups. The performance of ZFS and SUS are better than random selection, but worse than the optimal solution because they are both heuristic in nature. We also observe that the system throughput increases with the number of users in the network for the Blossom algorithm, ZFS, and SUS. This is because the search space for each algorithm becomes larger so there is a higher chance to find a better grouping result.

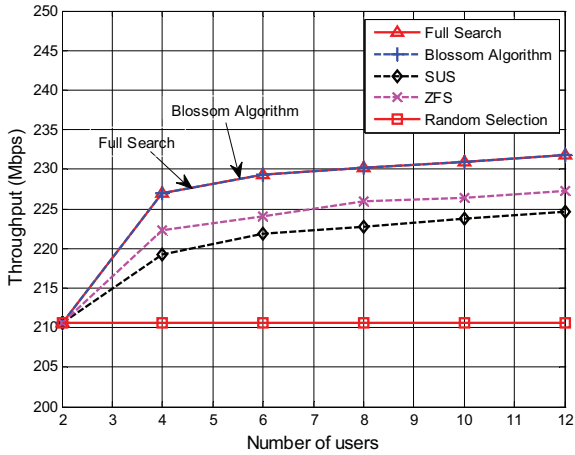

**Fig. 4.** System throughput as a function of the total number of the users in the network for $N_u = 2$.

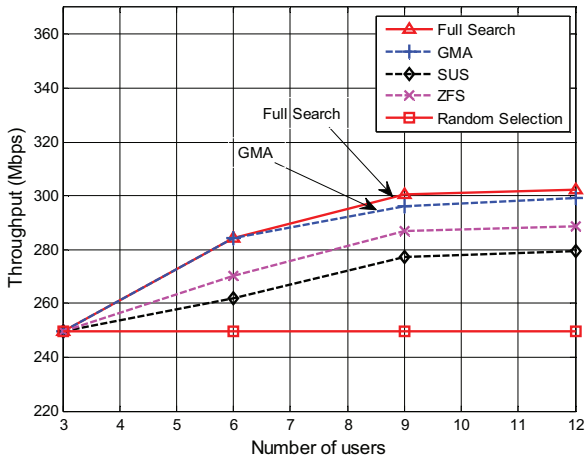

**Fig. 5.** System throughput as a function of the total number of the users in the network for $N_u = 3$.

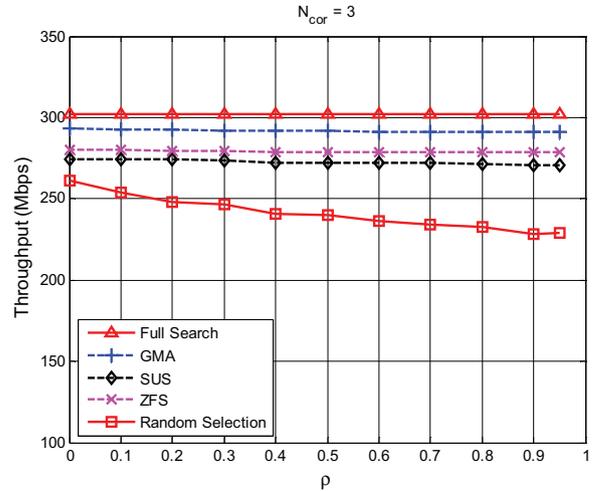

**Fig. 6.** Throughput as a function of $\rho$ when there are 3 correlated users in the network.

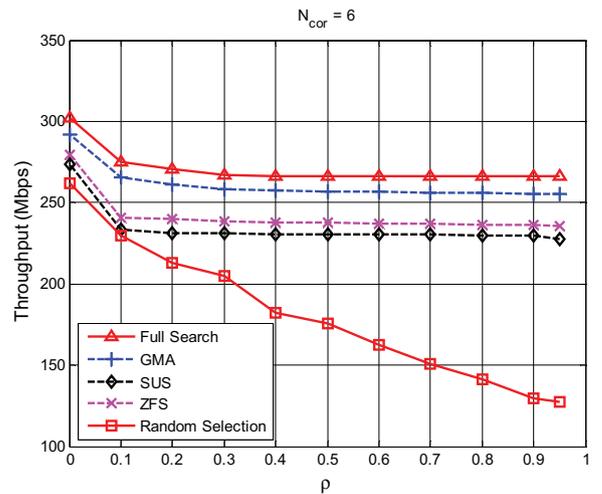

**Fig. 7.** Throughput as a function of $\rho$ when there are 6 correlated users in the network.

Fig. 5 shows the system throughput results for the scenario where up to 3 stations can be grouped for MU transmission. We observe that our proposed GMA is closest to the optimal solution which is bounded by full search. The proposed algorithm outperforms ZFS and SUS. The reason for this can be explained by the group splitting and regrouping mechanisms discussed in Section 3.2.2. This is to say, effective reshuffling of the users to different groups leads to high system throughput, which is very close to the optimum. From Fig. 5, we observe that our proposed algorithm leads to a system throughput of 98% of the optimal throughput for the considered scenario.

When we examine how users are grouped, we find that MU transmission is preferred by all approaches. That is to say, the more users can be grouped for transmission, the higher the system throughput becomes. This means that the correlation among the channels perceived at the different antennas is very low, which can be explained by the random placement of the receive antennas in the office. To mimic an outdoor environment where there is typically much higher channel correlation and line-of-sight (LOS) effects, we evaluate the algorithms using simulated correlated channels in Section 4.3.

### 4.3. Evaluation based on the simulated channels

We have observed that the all the methods attempt to group as many users as possible for transmission. The low channel correlation explains why even the random selection approach can lead to a reasonable system throughput. Now we consider the outdoor case where there is typically a higher correlation between users' channels and larger line-of-sight (LOS) effects than indoor environments. The channel is simulated using a Rician fading channel where the $k$ factor is set to 8 dB and the user number is 12.

Fig. 6 shows the system throughput for the different algorithms as a function of the correlation coefficient in the scenario where there are 3 correlated users. We observe that the throughput performance of the full search, the proposed GMA, ZFS, and SUS do not depend on the value of the correlation coefficient $\rho$. This is because $MU_3$ is always used as long as none of the three correlated users are grouped together. On the other hand, the performance of random selection degrades with increasing value of $\rho$. This is reasonable because the correlated users can be grouped together and higher correlation among users in the MU transmission leads to significantly lower data rates.

An even more severe throughput drop for the random selection is observed in Fig. 7, which shows the results for the case of

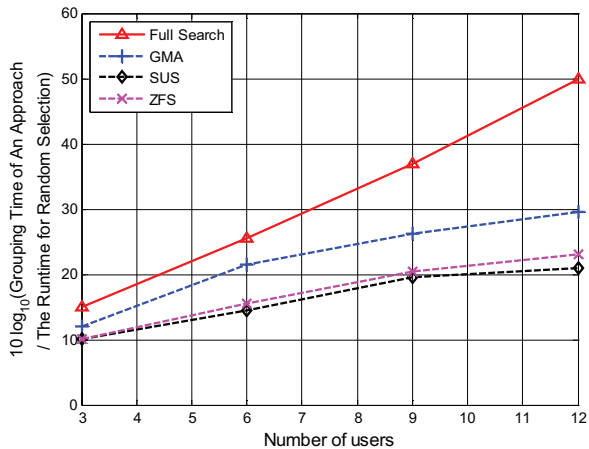

**Fig. 8.** Relative complexity with respect to random selection vs. network size.

6 correlated users in the network. For example, the system throughput is around 235 Mbps in the 3-correlated-user scenario if $\rho$ equals 0.6 and drops to 160 Mbps when the number of correlated users increases to six. This is because with six as opposed to three correlated users, there is a higher chance that the random selection would group correlated users together for MU transmission. Fig. 7 clearly indicates the benefit of flexible group sizes. The throughput performances of all approaches degrade as the value of channel correlation index $\rho$ increases. Random selection always chooses maximum group sizes, which leads to a considerably higher relative throughput loss than other approaches for $\rho \geq 0.2$. This is so because GMA, SUS, and ZFS avoid grouping the correlated user together and, thus, $MU_2$ or even $SU$ transmission may be preferred over $MU_3$. The degradation from $MU_3$ to $MU_2$ and/or $SU$ leads to the system throughput drops for GMA, SUS, and ZFS. Among all the analyzed algorithms, the proposed GMA performs closest to optimum and achieves at least 93% of the optimal system throughput.

We then measured the runtime of grouping functionality among all the approaches. To compare the complexity of the approaches, we set the random selection as the baseline. Fig. 8 plots the relative complexity with respect to random selection in dB.

As expected, full search needs considerably more time to complete and is not a feasible solution when the network size increases. Although GMA requires more time than SUS and ZFS, all the heuristic algorithms' runtime increases with the network size much slower than full search as they are all polynomial time solutions. Notice that the Blossom and Hungarian algorithms can be parallelized as investigated in [32,33]. Using Application-Specific Integrated Circuits (ASICs) for hardware acceleration, the runtime can be further reduced.

In this paper, similar to [14,15,19–21], we set the AP to acquire CSI of all active users before performing downlink MU transmission in a TXOP period. The focus of this paper is to provide a low-complexity grouping algorithm to effectively select user(s) for single user transmit beamforming or multi-user MIMO transmission. The proposed grouping strategy works with any sounding interval and is independent of how the sounding interval is decided. Large-scale networks, with large sounding overhead which are beyond the scope of this paper, can also be divided into a few small scale network sets to be served in turn and apply the proposed grouping strategy.

## 5. Conclusion

This paper formulates the optimization problem for choosing among different MU-MIMO transmission modes and SU-TxBF for practical WLAN deployments in emerging indoor-outdoor scenarios. A hypergraph matching-based solution is proposed and demonstrated to be effective for solving the user grouping problem that maximizes throughput while providing MU-MIMO air time fairness. The optimal grouping result is first derived for the 2-user group case, followed by an extension to the more general case where any number of users can be assigned to groups of different sizes as a function of the radio environment. A computationally efficient approach is proposed for practical system implementation. Using our collected indoor channel measurements, we have evaluated the proposed algorithms under the specification framework of IEEE 802.11ac. The results demonstrate the suitability of the algorithms for low and high correlated channels, emulating indoor and outdoor practical WLAN deployments. Our proposed algorithm achieves at least 93% of the optimal system throughput in all test cases and is within 98% of the optimal solution for the conducted indoor experiments. We are currently extending our work to analyze the system delay in scenarios where users have different traffic patterns and queuing lengths. The design and implementation of the parallelized version of the proposed algorithm for ASIC design is also planned for our future work. In addition, future work will jointly analyze the user grouping and sounding problem as well as the implications of large-scale networks.

## Acknowledgments

This work was partially supported by Qualcomm Atheros and Wireless@Virginia Tech.

## References

[1] Juniper Research. http://www.juniperresearch.com/press/press-releases/wifi-to-carry-60pc-of-mobile-data-traffic-by-2019, (accessed 23-Sep)-2015.
[2] E. Perahia, R. Stacey, Next generation wireless LANs: 802.11n and 802.11ac, Cambridge University Press, 2013.
[3] M.X. Gong, B. Hart, S. Mao, Advanced wireless LAN technologies: IEEE 802.11ac and beyond, GetMobile Mob. Comput. Commun. 18 (2015) 48–52.
[4] U.C. Lai, R.D. Murch, A transmit preprocessing technique for multiuser MIMO systems using a decomposition approach, IEEE Trans. Wireless Commun. 3 (2004) 20–24.
[5] C. Windpassinger, R.F.H. Fischer, T. Vencel, J.B. Huber, Precoding in multiantenna and multiuser communications, IEEE Trans. Wireless Commun. 3 (2004) 1305–1316.
[6] A. Goldsmith, S.A. Jafar, N. Jindal, S. Vishwanath, Capacity limits of MIMO channels, IEEE J. Sel. Areas Commun. 21 (2003) 684–702.
[7] G. Caire, S. Shamai, On the achievable throughput of a multiantenna Gaussian broadcast channel, IEEE Trans. Inf. Theory 49 (2003) 1691–1706.
[8] E. Aryafar, N. Anand, T. Salonidis, E.W. Knightly, Design and experimental evaluation of multi-user beamforming in wireless LANs, in: Proceedings of the sixteenth Annual International Conference on Mobile Computing and Networking, 2010, pp. 197–208.
[9] H. Yu, L. Zhong, A. Sabharwal, D. Kao, Beamforming on mobile devices: a first study, in: Proceedings of the 17th Annual International Conference on Mobile Computing and Networking, 2011.
[10] W. Shen, K. Lin, S. Gollakota, M. Chen, Rate adaptation for 802.11 multiuser MIMO networks, IEEE Trans. Mob. Comput. 13 (2014) 35–47.
[11] Y. Qian, B. Fan, K. Zheng, Group-based user pairing for virtual MIMO in LTE, J. China Univ. Posts Telecommun. 14 (2007) 38–42.
[12] A. Hottinen, E. Viterbo, Optimal user pairing in downlink MU-MIMO with transmit precoding, in: 6th International Symposium on Modeling and Optimization in Mobile, Ad Hoc, and Wireless Networks and Workshops, 2008, pp. 97–99.
[13] C.H. Papadimitriou, K. Steiglitz, Combinatorial optimization: algorithms and complexity, Courier Corporation, 1998.
[14] T. Yoo, A. Goldsmith, On the optimality of multiantenna broadcast scheduling using zero-forcing beamforming, IEEE J. Sel. Areas Commun. 24 (2006) 528–541.
[15] G. Dimic, N.D. Sidiropoulos, On downlink beamforming with greedy user selection: performance analysis and a simple new algorithm, IEEE Trans. Signal Process. 53 (2005) 3857–3868.
[16] T. Kuo, K. Lee, K.C. Lin, M. Tsai, Leader-contention-based user matching for 802.11 multiuser MIMO networks, IEEE Trans. Wireless Commun. 13 (2014) 4389–4400.
[17] F. Kaltenberger, D. Gesbert, R. Knopp, M. Kountouris, Correlation and capacity of measured multi-user MIMO channels, in: IEEE 19th International Symposium on Personal, Indoor and Mobile Radio Communications, 2008, pp. 1–5.

[18] V. Jones, H. Sampath, Emerging technologies for WLAN, IEEE Commun. Mag. 53 (2015) 141–149.
[19] E. Kartsakli, N. Zorba, L. Alonso, C. Verikoukis, Multiuser MAC protocols for 802.11n wireless networks, in: IEEE International Conference on Communications, 2009, pp. 1–5.
[20] R. Liao, B. Bellalta, C. Cano, M. Oliver, DCF/DSDMA: enhanced DCF with SDMA downlink transmissions for WLANs, in: Baltic Congress on Future Internet Communications, 2011, pp. 96–102.
[21] Z. Zhang, S. Bronson, J. Xie, H. Wei, Employing the one-sender-multiple-receiver technique in wireless LANs, in: IEEE INFOCOM, 2010, pp. 1–9.
[22] M. Heusse, F. Rousseau, G. Berger-Sabbatel, A. Duda, Performance anomaly of 802.11b, in: IEEE INFOCOM, 2003, pp. 836–843.
[23] G. Tan, J. Guttag, Time-based fairness improves performance in multi-rate WLANs, USENIX Annual Technical Conference (2004).
[24] L. Jiang, S. Liew, Proportional fairness in wireless LANs and ad hoc networks, in: IEEE Wireless Communications and Networking Conference, 3, 2005, pp. 1551–1556.
[25] R.V. Rasmussen, M.A. Trick, Round robin scheduling – a survey, Eur. J. Oper. Res. 188 (2008) 617–636.
[26] C. Berge, E. Minieka, Graphs and Hypergraphs, North-Holland Publishing Company, 1973.
[27] Y.H. Chan, L.C. Lau, On linear and semidefinite programming relaxations for hypergraph matching, Math. Program. 135 (2012) 123–148.
[28] E. Angel, A survey of approximation results for local search algorithms, in: Efficient Approximation and Online Algorithms, Springer, 2006, pp. 30–73.
[29] P. Berman, A d/2 approximation for maximum weight independent set in d-claw free graphs, in: Algorithm Theory-SWAT 2000, Springer, 2000, pp. 214–219.
[30] H.W. Kuhn, The Hungarian method for the assignment problem, Nav. Res. Log. Q. 2 (1955) 83–97.
[31] V. Kolmogorov, Blossom V: a new implementation of a minimum cost perfect matching algorithm, Math. Program. Comput. 1 (2009) 43–67.
[32] C.N.K. Osiakwan, S.G. Akl, A perfect speedup parallel algorithm for the assignment problem on complete weighted bipartite graphs, in: International Conference on Databases, Parallel Architectures and Their Applications, 1990, pp. 293–301.
[33] S. Ahn, S. Park, M. Chertkov, J. Shin, Minimum weight perfect matching via blossom belief propagation, in: Advances in Neural Information Processing Systems, 2015, pp. 1288–1296.

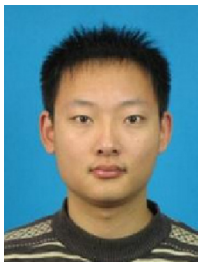

**Xiaofu Ma** received the B.S. degree in electronics from Northwest University, Xi'an, China, in 2008, and the M.S. degree in computer science from Tongji University, Shanghai, in 2011, and is currently pursuing the Ph.D. degree in electrical engineering at Virginia Polytechnic Institute and State University, Blacksburg, USA. His research interests include network protocol and optimization, cognitive radio networks, dynamic spectrum access, spectrum sharing, chipless RFID, wireless healthcare and wireless local area networks.

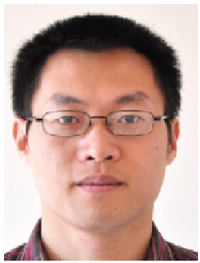

**Qinghai Gao** received his B.S. and M.S. from Xidian University, Xi'an, China, in 1999 and 2002 respectively, and the Ph.D. degree from Arizona State University in 2008. He is currently with Qualcomm Atheros, San Jose, CA. His research interests are in the areas of cross-layer optimization and cooperative communications in wireless networks.

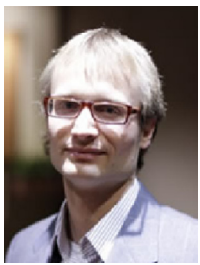

**Vuk Marojevic** received the Dipl.-Ing. degree from the University of Hannover in 2003 and the Ph.D. degree from the Universitat Politècnica de Catalunya (UPC), Spain, in 2009, both in electrical engineering. He is currently a research associate at Virginia Polytechnic Institute and State University, Blacksburg, USA. His research interests include software-defined and cognitive radios and wireless communications infrastructure and waveforms for reliable high-capacity networks.

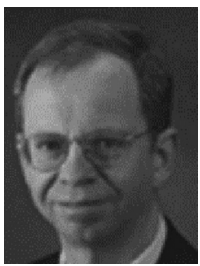

**Jeffrey H. Reed** received the B.S.E.E., M.S.E.E., and Ph.D. degrees from the University of California Davis, Davis, CA, USA, in 1979, 1980, and 1987, respectively. He is the founder of Wireless @ Virginia Tech, and served as its director until 2014. He is the Founding Faculty member of the Ted and Karyn Hume Center for National Security and Technology and served as its interim Director when founded in 2010. He is cofounder of Cognitive Radio Technologies (CRT), a company commercializing of the cognitive radio technologies; Allied Communications, a company developing technologies for 5 G systems; and for Power Fingerprinting, a company specializing in security for embedded systems. Dr. Reed became Fellow to the IEEE for contributions to software radio and communications signal processing and for leadership in engineering education, in 2005. Dr. Reed is a Distinguished Lecturer for the IEEE Vehicular Technology Society. In 2013, he was awarded the International Achievement Award by the Wireless Innovations Forum. In 2012, Dr. Reed served on the President's Council of Advisors of Science and Technology Working Group that examines ways to transition federal spectrum to allow commercial use and improve economic activity.